\documentclass[twocolumn,superscriptaddress,floatfix,prl,showpacs]{revtex4}
\usepackage[dvips]{graphicx}

\begin{document}
\title{Charge dynamics of doped holes in high T$_c$ cuprates 
- A clue from optical conductivity}
\author{A.~S.~Mishchenko}
\affiliation{Cross-Correlated Materials Research Group, RIKEN,
2-1 Hirosawa, Wako, Saitama, 351-0198, Japan}
\affiliation{RRC ``Kurchatov Institute", 123182, Moscow, Russia}

\author{N.~Nagaosa}
\affiliation{Cross-Correlated Materials Research Group, RIKEN,
2-1 Hirosawa, Wako, Saitama, 351-0198, Japan}
\affiliation{CREST, Department of Applied Physics, The University of Tokyo,
7-3-1 Hongo, Bunkyo-ku, Tokyo 113, Japan} 

\author{Z.-X.~Shen}
\affiliation{Department of Physics, Applied Physics, and Stanford Synchrotron 
Radiation Laboratory, Stanford University, Stanford, CA 94305 USA}
 
\author{G.~De~Filippis}
\affiliation{Coherentia-CNR-INFM and Dip. di Scienze Fisiche - 
Universit\`{a} di Napoli Federico II - I-80126 Napoli, Italy}

\author{V.~Cataudella}
\affiliation{Coherentia-CNR-INFM and Dip. di Scienze Fisiche - 
Universit\`{a} di Napoli Federico II - I-80126 Napoli, Italy}

\author{T.~P.~Devereaux}
\affiliation{Department of Physics, University of Waterloo, Ontario, 
Canada N2L 3GI}
\affiliation{Department of Physics, Applied Physics, and Stanford Synchrotron 
Radiation Laboratory, Stanford University, Stanford, CA 94305 USA}
  
\author{C.~Bernhard}
\affiliation{Department of Physics and Fribourg Center for Nanomaterials, 
University of Fribourg, CH-1700 Fribourg, Switzerland}

\author{K.~W.~Kim}
\affiliation{Department of Physics and Fribourg Center for Nanomaterials, 
University of Fribourg, CH-1700 Fribourg, Switzerland}

\author{J.~Zaanen}
\affiliation{Instituut Lorentz for Theoretical Physics, Leiden University,  
POB 9506, 2300 RA Leiden, The Netherlands}

\begin{abstract}
{The charge dynamics in weakly hole doped high temperature superconductors 
is studied in terms of the accurate numerical solution to a model of 
a single hole interacting with a quantum lattice in an antiferromagnetic 
background, and accurate far-infrared ellipsometry measurements. 
The experimentally observed two electronic bands in the infrared spectrum 
can be identified in terms of the interplay between the electron 
correlation and electron-phonon interaction resolving the long standing 
mystery of the mid-infrared band.}
\end{abstract}


\maketitle
It is now widely recognized that the physics of doping holes into a Mott 
insulator is the key concept to understand the high temperature 
superconductivity in cuprates \cite{LeeNagWen}. 
An appealing scenario is that the spin singlet
pairs already existing in the insulating antiferromagnet turn into the
superconducting Cooper pairs when the doped holes introduce the charge 
degrees of freedom. On the other hand, it is also noted that the parent 
compound is an ionic insulator, where the polar electron-phonon  
interaction (EPI) plays an essential role, and it is expected that this 
strong EPI continues to be of quite vital importance even at finite doping. 
Therefore, the quantum dynamics of the doped holes is essentially influenced 
by both the magnetic fluctuations and quantum phonons.

It has been recognized that 
the charge dynamics is determined not by the large Fermi surface but by the 
doped holes in the underdoped region \cite{LeeNagWen}. 
Therefore, it is a reasonable approach to 
consider the charge-current dynamics of the holes (not electrons) interacting 
with the quantum phonons and magnons simultaneously to analyze the infrared 
optical conductivity (OC).
The basic features of the observed d.c. conductivity and OC follow 
Ref.~\cite{BasTim,Kastner,ManoRev94}.
The OC $\sigma(\omega)$ in 
undoped material reveals the charge transfer band at $\omega \cong 1.5$eV 
between the p-orbitals of oxygen and d-orbitals of copper. With 
doping, a low energy part of $\sigma(\omega)$ develops revealing the 
dynamics of the doped holes as a function of the frequency $\omega$. 
In particular, the Drude weight is shown to be proportional to $x$ 
even in the absence of antiferromagnetic long range order, 
while the relaxation rate $1/\tau$ 
is proportional to the temperature $T$. 
Therefore, this dependence of the Drude weight finds a  natural 
explanation as directly reflecting the hole concentration, as mentioned above, 
while the $T$-dependence is due to the lifetime of the holes. 
The higher frequency 
$\sigma(\omega)$, on the other hand, has not been well-understood. 
Especially the mid-infrared (MIR) peak with dependent on doping energy
at around $\omega_{\mbox{\scriptsize MIR}} \cong 0.5$eV 
is still controversial \cite{LeeBas2005}, with interpretations involving an 
 $\omega$-dependence of $1/\tau$, transitions between the Zhang-Rice 
singlet state to the upper Hubbard band, and magnon sidebands.

For this problem, angle-resolved-photoemission spectroscopy (ARPES) offers 
an important clue. ARPES in undoped parent compounds measures the 
spectral function of a single hole left behind 
when an electron is kicked out from the sample by the incident light 
\cite{Shen_03, Kyle_Dop}. 
The corresponding theoretical analysis has pointed out the role of 
the composite polaronic effect due to electron-magnon and electron-phonon 
(el-ph) couplings \cite{tJpho,Rosch}. Therefore, it is expected that these 
two interactions are indispensable to understand the infrared optical 
spectra as well.

\begin{figure}[bht]
\begin{center}
\includegraphics[width=8cm]{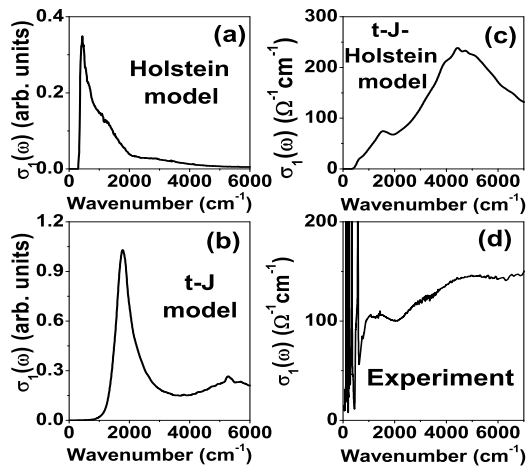}
\end{center}
\caption{\label{fig:fig1} Comparison of typical OCs of different 
models in 2D with experimental data of heavily underdoped cuprates:
(a) Holstein model at $\lambda=0.44$; (b) t-J model at $J=0.3$;
(c) t-J-Holstein model for $J=0.3$ and $\lambda=0.39$; 
(d) in-plane OC of 1.5\% hole doped 
(Eu$_{1-x}$Ca$_x$)Ba$_2$Cu$_3$O$_6$  at $T=10$K.
The energy dependence of the theoretical data is presented in wavenumbers 
assuming $t=0.3$eV (1eV=8065.5cm$^{-1}$). 
The absolute value of the theoretical $\sigma_1$ 
is evaluated using the experimental hopping distance $a=3.86\AA$ and
bulk hole concentration $n_h=1.72 \times 10^{-23}$ cm$^{-3}$.    
} 
\end{figure}

We study in this paper theoretically the OC of the 
single-hole doped into a Mott insulator described by the t-J-Holstein model, 
and its numerical solution in terms of the Diagrammatic  
Monte Carlo (DMC) simulation. 
Previously, the OC of the t-J-Holstein model has been calculated by 
exact diagonalization of small clusters \cite{Fehske98}, in the 
non-crossing approximation NCA for both 
magnetic and lattice variables \cite{KyungMukhin96}, and for the case 
of infinite dimension \cite{Cappelluti}. 
Compared with these approximate methods, our DMC simulation provides 
more accurate solution for the infinite system without approximation 
associated with the phonon sector \cite{MPSS} and for proper lattice 
geometry and dimension.
The only approximation is the NCA for magnons which is shown to be 
sufficiently good for the parameters considered below \cite{Liu_92,RoGu_SCBA}.
These results are compared with accurate far-infrared 
ellipsometry measurement as well as with previously published data.
The infrared ellipsometry measurements have been performed with a home-built 
ellipsometer attached to a Bruker Fast-Fourier spectrometer at the IR beamline 
of the ANKA synchrotron at FZ Karlsruhe, D at 70-700 cm$^{-1}$ and with 
conventional light sources at 500-7000 cm$^{-1}$ \cite{Bern1}.

Figure~\ref{fig:fig1} summarizes our main results, where four panels 
for the infrared parts of $\sigma(\omega)$ are compared. 
Figures~\ref{fig:fig1}(a), (b) and (c) show the calculated  $\sigma(\omega)$ 
for the Holstein model (el-ph coupling only), t-J model (el-magnon only),
and the t-J-Holstein model (el-ph and el-magnon couplings), respectively, while
Fig.~\ref{fig:fig1}(d) presents the experimental observation. 
Neither the t-J model \ref{fig:fig1}(b), nor the Holstein model 
\ref{fig:fig1}(a) bear resemblance to the experiment \ref{fig:fig1}(d), while 
their combination \ref{fig:fig1}(c) at least qualitatively 
reproduces the salient experimental features.
The clear signature of the 
experiment is that there are two prominent electronic components, i.e., 
the so-called MIR band at around 
$\omega_{\mbox{\scriptsize MIR}}= 4600$cm$^{-1}$ and the 
lower energy one at around $\omega= 1000$cm$^{-1}$ that is located just above 
the infrared active phonon modes which show up as sharp peaks below 
$800$cm$^{-1}$. 
The lower energy peak roughly corresponds to that seen in 
Fig.~\ref{fig:fig1}(c) due to the phonon side-band. 
However, the MIR peak does not correspond to that of t-J model which 
occurs at around  
$\omega_{\mbox{\scriptsize t-J}} \cong 2J \cong 2000$cm$^{-1}$.
Instead of that, according with experiment, it is shifted to higher
energies.

As it can be seen from the results below, the coupling to two kinds 
of bosonic excitations results in two separate peaks because of the  
essentially different nature of the electron-magnon and el-ph couplings 
\cite{T-depend} and significantly different energy scales of the magnetic and 
lattice excitations which are involved in the optical transitions. 
The magnons with large characteristic energy $\sim 2J$ are weakly bound 
to the hole. 
To the contrary, the phonons are adiabatic and the EPI is considerable.
As shown below, the lower energy peak is the phonon sideband with the 
threshold at the phonon energy \cite{Optics} and the higher energy peak
is the magnon sideband of the lower peak.     
The reason for the apparent shift of the $2J$ peak to higher energies
is most evident in the strong-coupling limit where the Franck-Condon picture 
for optical processes is valid \cite{Optics}
and the fluctuations of energies of different lattice sites, with the 
characteristic scale of Franck-Condon energy, can be considered as being 
frozen. 
Then,  the energy cost of the transition of the hole from the ground 
state to excited states of the t-J model with frozen lattice is the sum 
of the energy of the emitted magnon and the Franck-Condon energy.
Hence, the two peaks in the OC are the consequence of the importance of both 
el-ph and magnetic interactions. 
The same information is encoded in a different way in the single-particle 
spectral function observed in ARPES, where the low energy 
quasiparticle peak of t-J model is affected by el-ph interaction. 
This low energy peak is separated into the broad 
Franck-Condon peak mimicking the  dispersion of the t-J model 
while the zero-phonon line with very small weight has almost no dispersion 
\cite{tJpho,Kyle_Dop}.

In the standard spin-wave approximation for the t-J model \cite{kane,Liu_92}, 
the dispersionless hole $\varepsilon_0 = const$ (annihilation operator is 
$h_{\bf k}$) propagates in the magnon (annihilation operator is $\alpha_{\bf 
k}$) bath
\begin{equation}
\hat{H}_{\mbox{\scriptsize t-J}}^{0} =
\sum_{\bf k} \varepsilon_0 h_{\bf k}^{\dagger} h_{\bf k} 
+ \sum_{\bf k} \omega_{\bf k} \alpha_{\bf k}^{\dagger} \alpha_{\bf k}
\label{h0}
\end{equation}
with magnon dispersion $\omega_{\bf k}=2J\sqrt{1-\gamma_{\bf k}^2}$, where
$\gamma_{\bf k}=(\cos k_x + \cos k_y) / 2$. The hole is scattered by magnons 
\begin{equation}
\hat{H}_{\mbox{\scriptsize t-J}}^{\mbox{\scriptsize h-m}} =
N^{-1/2} \sum_{\bf k , q} M_{\bf k , q} 
\left[ h_{\bf k}^{\dagger} h_{\bf k-q} \alpha_{\bf k} + h.c.
\right] 
\label{h-m}
\end{equation}  
with the standard scattering vertex $M_{\bf k , q}$ \cite{ManoRev94}. 

\begin{figure}[bht]
\begin{center}
\includegraphics[width=8cm]{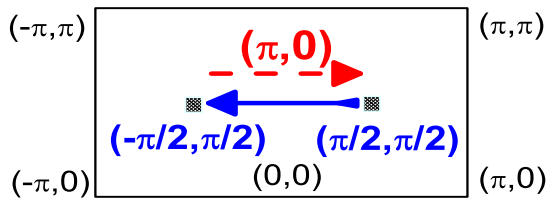}
\end{center}
\caption{\label{fig:fig2} (color online) 
Electronic transition (solid arrow) and emitted magnon (dashed arrow)
responsible for the $2J$ peak in the OC of the t-J model.
} 
\end{figure}

\begin{figure}[bht]
\begin{center}
\includegraphics[width=8cm]{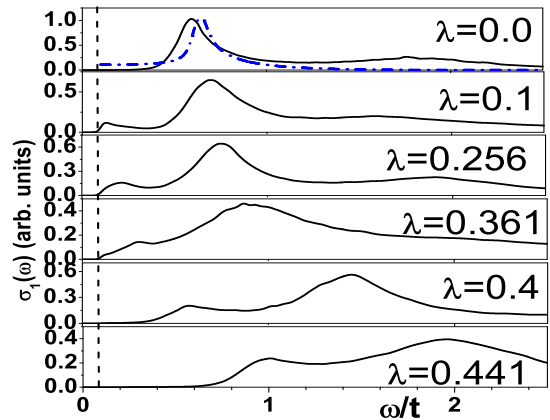}
\end{center}
\caption{\label{fig:fig3} (color online) 
OC of a single hole in the t-J-Holstein model
at $J/t=0.3$ with various EPI coupling constants 
$\lambda$. 
The vertical dashed line at $\omega/t=0.1$ indicates the phonon frequency. 
The dash dotted line in the upper panel is the result of Eq.~(\ref{GuVi}).
} 
\end{figure}

The OC of the t-J model has been calculated by 
various methods in numerous papers giving mutually consistent results,  
e.g.\ \cite{RiceZhang89,Eder96}. 
To confirm the validity of our method, we first reproduce the results 
for different J/t ratios \cite{Optics_why}. We find the well known peak  
at around $\omega_{\mbox{\scriptsize t-J}} \cong 2J$, the 
origin of which has not been settled yet. It is also difficult to identify 
this peak with the observed MIR peak since the energy of the latter is about 
2 times higher than $2J$. Another source for skepticism is the opposite 
doping dependence of the $\omega_{\mbox{\scriptsize t-J}}$ and 
$\omega_{\mbox{\scriptsize MIR}}$ energies \cite{Eder96}. 
Therefore, we conclude that the t-J model can not explain the 
observed OC even at very small dopings. 
Before introducing the EPI, let us first provide 
an interpretation of the $\omega_{\mbox{\scriptsize t-J}} \cong 2J$ peak. 
This peak comes from the hole excitations within the coherent band of the t-J 
model from the ground state at $(\pi/2,\pi/2)$ to the neighborhood of 
$(-\pi/2,\pi/2)$ point, assisted by emission 
of single magnon with energy $\approx 2J$ and a momentum $(\pi,0)$
(Fig.~\ref{fig:fig2}).
To prove this point we computed the OC by disentangling magnon and hole 
terms (exact at the lowest order in $t$) and then calculating the
correlation function involving the hole operators in the subspace with 
one magnon (See upper panel in fig.~\ref{fig:fig3})
\begin{eqnarray}
&\Re \sigma(\omega)=4 \pi t^2 e^2 ( \omega N )^{-1} \sum_{\vec{q}}
\left| \left\langle \psi^{(1)}_{\vec{k_0}-\vec{q}} \left | O_{\vec{q}} \right
 | \psi^{(1)}_{\vec{k_0}} \right\rangle \right| ^2 \times &
\nonumber
\\
&\delta \left[ \omega - \omega_q - (E^{(1)}_{\vec{k_0}-\vec{q}}-
E^{(1)}_{\vec{
k_0}}) \right],&
\label{GuVi}
\end{eqnarray}
Here $\left | \psi^{(1)}_{\vec{k}} \right\rangle $ is the lowest eigenstate
in the subspace with one magnon with energy 
$E^{(1)}_{\vec{k}}$, $\vec{k_0}=
(\pi/2,\pi/2)$ and 
$O_{\vec{q}}=\sum_{\vec{k}} h^{\dagger}_{\vec{k}-\vec{q}} h _{\vec{k}} 
C(\vec{
k}-\vec{q},\vec{k})$ \cite{about_c}.
By direct inspection of the sum over ${\bf q}$  in Eq.~\ref{GuVi} we found 
that the main contribution to the OC comes from the transfer of magnon 
momentums around $(\pi,0)$. It can be traced out from the 
${\bf q}$-dependence of the vertex 
$|C({\bf k}_0-{\bf q},{\bf k}_0)|^2$.

We now turn to the EPI added to the t-J model, the importance 
of which has been already established as mentioned above. When the model is 
updated to the t-J-Holstein model, the hole interacts with dispersionless 
(frequency $\Omega=\mbox{const}$) optical phonons by short range coupling 
$\gamma$   
\begin{equation}
\hat{H}^{\mbox{\scriptsize e-ph}} = 
\Omega \sum_{\bf k} b_{\bf k}^{\dagger} b_{\bf k}
+
\frac{ \gamma}{\sqrt{N}}  \sum_{\bf k , q}  
\left[ h_{\bf k}^{\dagger} h_{\bf k-q} b_{\bf k} + h.c.
\right] \;.
\label{e-ph}
\end{equation} 
In units of $t=1$ we parametrize the dimensionless EPI constant as 
$\lambda=\gamma^2 /(4t\Omega)$, choosing the value $J/t=0.3$, and setting the 
phonon frequency $\Omega=0.1$ \cite{omega}. 

\begin{figure}[t]
\begin{center}
\includegraphics[width=9cm]{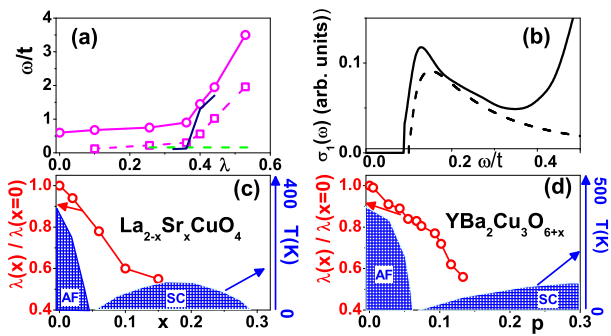}
\end{center}
\caption{\label{fig:fig4} (color online) (a)  
  EPI coupling dependence of the energies of the dominating peak  
(solid line with circles) and EPI-mediated feature (dashed line with squares)
of the OC. 
The energies of dominating phonon peak of the OC of the Holstein model at 
$t=1$ (dashed line) and those at $\tilde{t}=0.4$ (solid line). 
(b) OC of the t-J-Holstein model (solid line) and OC of the effective Holstein 
model with $\tilde{t}=0.4$ (dashed line) at the same coupling $\lambda=0.1$.  
Ratio of the effective EPI constant at doping $x$ (or real in-plane 
concentration $p$ for YBCO \cite{Hardy06}) to that 
at zero doping estimated from the MIR peak position for (c) LSCO 
and (d) YBCO mapped on the phase diagrams.
} 
\end{figure}

Figure~\ref{fig:fig3} shows the effect of EPI on the OC of a single hole in 
t-J-Holstein model. 
At weak EPI, an absorption starts and shows a peak right above the phonon 
frequency.  
This apparent two-peak structure of the MIR response of underdoped 
cuprates can be tacitly discerned from many previous measurement 
(Fig.~3 in \cite{Thomas1992} and Fig.~9 in  \cite{BasTim}) and 
is clearly seen from the low-temperature in-plane OC of 1.5\% hole doped 
Eu$_{1-x}$Ca$_x$Ba$_2$Cu$_3$O$_6$ measured by ellipsometry 
(Fig.~\ref{fig:fig1}(d)).  
This low energy EPI-mediated peak stays close 
to phonon energy up to the self-trapping transition point which is, for given 
parameters of the model, is located at $\lambda \approx 0.4$
\cite{tJpho}. Indeed, according to the dependence of the dominant $2J$ 
contribution and the low energy peak on $\lambda$ 
(Fig.~\ref{fig:fig4}a), the transition from the weak- to the strong-coupling 
regime occurs at this coupling strength.    
We note that both the t-J model and polaron physics is crucial to explain the 
very existence of the two-peak structure of the OC.

To understand the nature of the low energy peak induced by EPI, we 
did a calculation of OC for the Holstein model without hole-magnon interaction 
with reduced transfer ${\tilde t} = 0.4t$ mimicking the mass enhancement,
which reproduces the self-trapping point of t-J-Holstein model with $t=1$ 
As seen in Fig.~\ref{fig:fig4}b, this effective Holstein model reproduces 
remarkably well the shape of the low energy feature of OC for 
the t-J-Holstein model. 

Since the effective EPI decreases with doping \cite{Lanzara01,MiCond} the 
reason of the experimentally observed \cite{Uchida1991,LeeBas2005,Cooper1993}
MIR mode softening is the change of the EPI. Comparing the position 
of the MIR mode with results of the t-J-Holstein model we can give a rough
estimate of the renormalization of the effective EPI with doping.
First, since the self-trapping point of the realistic extended tt$'$t$''$-J
model is 
$\lambda_{\mbox{\scriptsize st}}^{\mbox{\scriptsize tt$'$t$''$-J}} 
\approx 0.6$ \cite{MNIsotope},
we scale the EPI strength in the Fig.~\ref{fig:fig4}a as 
$\lambda \to 1.5 \lambda$. 
Second, since the quadratic dependence of the energy scales of the OC 
response on $\lambda$ is a known property of the strong coupling 
regime \cite{Optics}, we extrapolate the OC data to larger EPI couplings. 
Using a quadratic extrapolation of the experimental 
data of the MIR peak in YBCO \cite{LeeBas2005,Cooper1993} and LSCO 
\cite{Uchida1991} to zero dopings, we arrive at the result  
in Fig.~\ref{fig:fig4}c--d which is 
in agreement with \cite{Lanzara01}. 
Moreover, since the analysis of the ARPES in undoped LSCO gives 
$\lambda \sim 1$ \cite{Rosch}, the data in Fig.~\ref{fig:fig4}c--d give 
absolute values of $\lambda^{\mbox{\scriptsize LSCO}}(x)$.
Finally, our result for $\lambda^{\mbox{\scriptsize LSCO}}(x)$ is in 
quantitative agreement with the values obtained from the ``kink'' 
angle in ARPES on LSCO \cite{MiCond}. 
Figures \ref{fig:fig4}c--d strongly suggest that superconductivity appears 
after the effective EPI decreases from strong to weak coupling, 
which liberates the coherent motion of the doped holes. The $x$-axis for 
YBCO should be translated to the hole doping concentration in the 
CuO$_2$-planes $p$, which makes the two phase diagrams Fig.~\ref{fig:fig4}(c) 
and (d) look almost similar. 
Therefore, the behavior of the effective EPI and phase diagram
seem almost universal in high-Tc cuprates.  However, the discussion above is 
restricted to the polaronic effect for the holes and not for the 
quasiparticles forming the large Fermi surface. 
The contribution of the EPI to the pairing 
hence is not excluded by the present analysis. 

Fruitful discussions with 
N.\ V.\ Prokof'ev, B.\ V.\ Svistunov and  G.\ A.\ Sawatzky
are acknowledged.
A.S.M.\ is supported by  RFBR 07-02-00067a. 
N.N. was partly supported by the
Grant-in-Aids from under the Grant No. 15104006, 16076205, and
17105002, and NAREGI Nanoscience Project from the Ministry of
Education, Culture, Sports, Science, and Technology, Japan.
C.B.\ and K.W.K.\ acknowledge technical support of Y.\ L.\ Mathias at the 
ANKA IR-beamline and funding by the Schweizer Nationalfonds (SNF)
through grant 200021-111690 and the Deutsche Forschungsgemeinschaft (DFG)
through grant BE 2684/1 in FOR538.


\begin{thebibliography}{99}
\bibitem{about_c} The vortex $C(\vec{k}-\vec{q},\vec{k})$ is defined in 
\cite{KyungMukhin96}.
\end{thebibliography}
\end{document}